\documentclass{aa} 
\usepackage{graphicx,psfig}

\newcommand{\Msun} {M$_\odot$}
\newcommand{\Lsun} {L$_\odot$}
\newcommand{\Teff} {T$_{\rm{eff}}$}
\newcommand{\Tstar} {T$_{\rm{eff}}$}
\newcommand{\Lstar} {L$_\star$}
\newcommand{\Mstar} {M$_\star$}
\newcommand{\Rstar} {R$_\star$}

\newcommand{\um} {$\mu$m}
\newcommand{\CQ} {CQ~Tau}
\newcommand{\Md} {M$_{d}$}
\newcommand{\Ri} {R$_{in}$}
\newcommand{\Rd} {R$_{d}$}

\newcommand{\degree} {$^{\rm o}$}

\newcommand{\simless}{\mathbin{\lower 3pt\hbox
      {$\rlap{\raise 5pt\hbox{$\char'074$}}\mathchar"7218$}}} 
\newcommand{\simgreat}{\mathbin{\lower 3pt\hbox
     {$\rlap{\raise 5pt\hbox{$\char'076$}}\mathchar"7218$}}} 

\begin{document}

\title{Large grains in the disk of CQ Tau}

\author{L. Testi\inst{1},
A. Natta\inst{1}, 
D. S.  Shepherd \inst{2},
D.J. Wilner \inst{3}
}

\institute{ 
    Osservatorio Astrofisico di Arcetri, INAF, Largo E.Fermi 5,
    I-50125 Firenze, Italy 
\and
National Radio Astronomy Observatory,P.O. Box O, Socorro, NM 87801
\and
Harvard-Smithsonian Center for Astrophysics, 60 Garden Street,
       Cambridge, MA 02138
}

\offprints{lt@arcetri.astro.it}
\date{Received ...; accepted ...}

\titlerunning{Large grains in the disk of CQ Tau}
\authorrunning {Testi et al.}

\abstract{
We present 7mm observations of the dusty disk surrounding the 10 Myr old
1.5 M$_\odot$ pre-main-sequence star CQ Tauri obtained at the Very Large Array
with 0.8 arcsecond resolution and 0.1 mJy rms sensitivity.
These observations resolve the 7mm emission in approximately the north-south
direction, confirming previous results obtained with lower resolution.
We use a two-layer flared disk model to interpret the observed fluxes
from 7mm to 1.3mm together with the resolved 7mm structure.
We find that the disk radius is constrained to the range 100 to 300 AU,
depending on the steepness of the disk surface density distribution.
The power law index of the dust opacity coefficient, $\beta$,
is constrained to be 0.5 to 0.7.
Since the models indicate that the disk is optically thin at millimeter
wavelengths for radii greater than 8 AU, the contribution of
an optically thick region to the emission is less than 10\%.
This implies that high optical depth or complex disk geometry cannot be
the cause of the observed shallow millimeter spectral index.
Instead, the new analysis supports the earlier suggestion that dust
particles in the disk have grown to sizes as large as a few centimeters.
The dust in the CQ Tauri system appears to be evolved much like that
in the TW Hydra system, a well-studied pre-main-sequence star of similar age
and lower mass.
The survival of gas-rich disks with incomplete grain evolution at such old ages
deserves further investigations.
}


\maketitle

\section{Introduction}

Grain growth from the sub-micron size, typical of dust in the ISM,
to millimeter
and centimeter sizes is the first step toward the formation of planets inside
circumstellar disks. We expect that such growth either occurs during the
pre-main--sequence (PMS) life of the star, when the circumstellar disks are
still gas-rich, or that it will never take place (Ruden~\cite{R99}).

So far, there is little  firm observational evidence for very large grains in
the disks of PMS stars (Beckwith et al.~\cite{Bea00}). The best indication
comes from the shallow wavelength dependence of the millimeter emission
from dust in most PMS disks (e.g. Beckwith \& Sargent~\cite{BS91}).
Namely, if the emission is optically thin,  the measured
spectral index $\alpha$ ($F_\nu \propto \nu^{\alpha})$ is directly related to
the dependence of the opacity on wavelength.  The grain size can then be
inferred from the opacity law, as long as optical depth and geometrical effects
can be sorted out.  This  requires that the spectral energy distribution (SED)
is well known and that the disk is spatially resolved at one 
or more wavelengths in the millimeter range (see e.g. the
discussion in Testi et al.~\cite{Tea01}).

Recently, this method has been applied to the nearby T Tauri star (TTS) TW~Hya
by Calvet et al.~(\cite{Cea02}), who have analyzed the SED of TW~Hya together
with  a 7~mm VLA map obtained by Wilner et al.~(\cite{Wea00}).  Using
self-consistent disk models, they showed that the observations could only be
reproduced if the grains in the outer disk have  grown  to centimeter sizes.

We report in this paper similar results on the PMS star \CQ. \CQ\ is a well
studied variable star of spectral type $\sim$A8, mass 1.5 \Msun, age
$\simgreat$10 Myr.  With a distance of about 100 pc (Hipparcos), it is one of
the closest young intermediate mass stars still surrounded by a massive
circumstellar disk (Natta et al.~\cite{NGM00}).  Its SED has been modelled as
due to a circumstellar disk by various authors (Chiang et al.~\cite{Cea01};
Natta et al.~\cite{Nea01}).  At millimeter wavelengths, it has been observed
with the OVRO interferometer (Mannings \& Sargent~\cite{MS97}; \cite{MS00}), with
Plateau de Bure (Dutrey et al. private communication; quoted by Natta et
al.~\cite{Nea01}) and with the VLA at 7mm and 3.6cm (Testi et al.~\cite{Tea01}).
The millimeter SED is characterized by a spectral index $\alpha_{mm}\sim 2.4-2.6$,
which suggests the presence of grains much larger than in the ISM. However,
as discussed by Testi et al. (2001), given the available data, two classes of
models can explain the observed fluxes and spectral index: small optically thick
disks with standard ISM dust, or bigger, optically thin disks with large grains.
As the disk was possibly resolved in our VLA D-array observations, the large
grains hypothesis was favoured. However, to solve the ambiguity between grains
and disk properties it is necessary to clearly resolve the disk and to derive
disk structural parameters (mainly the outer radius and surface density
distribution). We have thus obtained new VLA 7mm observations from a more
extended VLA configuration to better resolve the disk, and
we model these new data together with previous millimeter-wave
observations.


\section {VLA 7~mm observations and results}

CQ~Tauri was observed at Q-band (7~mm, 43~GHz) with the NRAO\footnote{The
National Radio Astronomy Observatory is a facility of the National Science
Foundation operated under cooperative agreement by Associated Universities,
INC.} VLA on July 21 and 31, 2001. The array was in the C configuration, and
the 24 antennas equipped with Q-band receivers offered baselines in the range
from 35~m to 3.4~km. 3C286 was used as primary flux calibrator, while
0547$+$273, whose flux was measured to be 0.52~Jy at 7~mm, was used as phase
and amplitude calibrator. Hourly pointing at X-band (3.6~cm, 8.4~GHz) was used
to ensure accurate pointing during the observing run.  The raw visibility data
was edited and calibrated within the AIPS software package using standard
techniques. The calibrated C-array ($u,v$) dataset was merged with the
calibrated D-array data from Testi et al.~(\cite{Tea01}) and then exported to
the MIRIAD package for deconvolution and modeling.  All maps presented in this
paper have been obtained with natural weighting of the visibilities; the
resulting synthesised beam FWHM is 0$\farcs$97$\times$0$\farcs$75 with position
angle $-$36$^\circ$, the noise level in the map is 0.1~mJy/beam.

\begin{figure}
\begin{center}
\leavevmode
\centerline{ \psfig{file=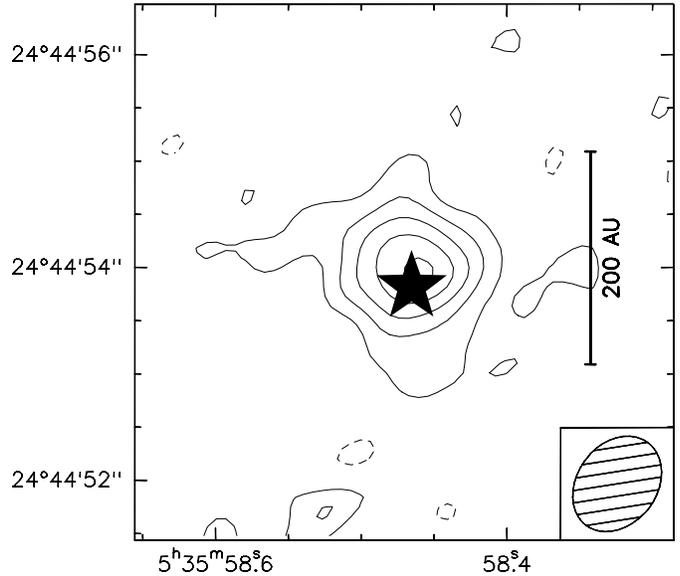,width=8.8cm,angle=-90} }
\end{center}
\caption{VLA C+D array 7~mm map of CQ~Tau. Contour levels start at 2$\sigma$
and are spaced by 2$\sigma$ (both negative and positive),
1$\sigma$=0.1~mJy/beam. The synthesised beam FWHM
(0$\farcs$97$\times$0$\farcs$75 at pa=$-$36$^\circ$) is shown by the shaded 
ellipse in the inset. The vertical bar shows the angular extent of 200~AU 
at the distance of CQ~Tau (100~pc). Based on Hipparchos position and proper
motion measurements, the star marks the position of the 
optical star at the epoch 2001.0, intermediate between the two VLA epochs.
The maximum proper motion shift between the two epochs is expected to be 
$\sim$28~mas in the north-south direction.
}
\label{fcqtau}
\end{figure}

The combined C- and D-array 7~mm map of CQ~Tau is shown in Figure~\ref{fcqtau}.
The new data confirm and extend the findings of Testi et al.~(\cite{Tea01}).
The millimeter peak is coincident with the position of the optical star as
measured from Hipparchos. The 7~mm emission fom the CQ~Tau system is resolved
and elongated approximately in the north-south direction. The 2$\sigma$ contour
has a maximum elongation slightly exceeding 2$^{\prime\prime}$ (200~AU). The
east-west ``plume'' at $\alpha\sim$5:35:58.55, $\delta\sim$24:44:54 is not
significant in our data and will be treated as a noise peak in this paper;
nevertheless it would be interesting to check in the future with additional
observations whether this is a real feature or not. Results from a two
dimensional gaussian fit to the continuum image are consistent with our
previous D-array resuls, albeit yelding a marginally smaller source size: the
deconvolved FWHM is 190$\times$70~AU.
Assuming a circularly symmetric source, the observed aspect ratio implies
an inclination of the disk axis from the line of
sight of $\sim$70$^\circ$, consistent within uncertainties with both our
D-array estimate and the 66\degree\ inclination derived from an analysis of the
optical photometric and polarimetric variability  by Natta \&
Whitney~(\cite{NW00}). The fact that with the new higher angular resolution
C-array data we derive deconvolved disk sizes smaller than those previously
determined with D-array data is not surprising. Due to the centrally peaked
nature of the disk emission, interferometric continuum maps
will always emphasize the
central regions of the disk and may loose the outer regions due to insufficient
surface brightness sensitivity or due to spatial filtering of the largest
scales (see also the discussion in Dutrey et al.~\cite{Dea96} and Wilner et
al.~\cite{Wea00}).

In order to correctly derive the disk parameters, the observed interferometric
maps should be compared with models of the disk emission sampled on the same
($u,v$) points and deconvolved in the same manner as the observations. This
analysis will be presented in the following section.

\section{Models}

We compare our observations to the predictions of disk
models based on the two-layer approximation first described by Chiang \&
Goldreich~(\cite{CG97}; CG97). The disk consists of a surface layer which
intercepts the stellar radiation and re-emits 1/2 of it toward the observer and
1/2 toward the disk midplane. The midplane is in turn heated by the radiation
reprocessed by the surface layer, and it is therefore cooler than grains on the
surface.  These models solve  the radiation transfer in a simple way, and allow
a self-consistent computation of the geometry of the disk, assuming hydrostatic
equilibrium in the vertical direction.  

Our models are an improved version of the original CG97 models. We have
implemented a more accurate  iterative scheme to compute the disk flaring angle
at each radius (Chiang et al.~\cite{Cea01}) and a more realistic dust model for
the surface grains.  In addition, we take advantage of the capability of the
two-layer scheme to use two different dust prescription for the midplane and
surface.  Details can be found in Natta et al.~(\cite{Nea01}) and Dullemond et
al.~(\cite{DDN01}).

At mm wavelengths, the predictions of these two-layer models are rather
robust, both in terms of integrated flux and of intensity profiles. For
example, there is no significant difference between the results from
our models and those from models where
the radiation transfer is properly solved in the vertical direction (Dullemond
et al.~\cite{DZN02}, Dullemond \& Natta~\cite{DN03a}). 
Furthermore, the use of a mean opacity for the grains in the
disk midplane does not affect the results in any relevant way (see Wolf \cite{W02} and references therein), and the effect of scattering of the
stellar radiation by grains in the disk (which is not included in the models)
is also negligible (Dullemond \& Natta~\cite{DN03b}).  

To compute a model, we need to specify the stellar properties (luminosity
\Lstar, effective temperature \Tstar, mass \Mstar\ and distance $D$) and some
disk parameters, namely the inner and outer radius (\Ri\ and \Rd, respectively)
and the surface density distribution $\Sigma=\Sigma_0 (R/R_0)^{-p}$, where $R_0$
is a fiducial radius.  The disk mass \Md\ is then fixed.  The other parameters
needed  are the dust properties on the surface and in the midplane. 
We describe the midplane dust opacity as a power-law
function of $\lambda$, $\kappa=\kappa_0 (\lambda/\lambda_0)^{-\beta}$, with
$\lambda_0=1.3$ mm and $\kappa_0$=1.0 cm$^2$ g$^{-1}$ of dust
(Hildebrand~\cite{H83}).
For grains
on the surface, we use a MRN mixture of graphite and silicates with 10\% of
cosmic carbon in grains of radius between 0.02 and 0.35 \um\ and 100\% cosmic
silicon in silicates of radius between 0.02 and 0.7 \um\ (see Natta et
al.~\cite{Nea01}).  This choice is  not important for the disk
properties at millimeter wavelengths, as long as the disk is optically thick to
the stellar radiation.

We assume that the stellar properties are known (\Lstar=4 \Lsun, \Teff=6900 K).
The exact value of the disk inner radius is not important, and we fix it at
\Ri=10 \Rstar, i.e., close to the dust sublimation radius.  We then vary
$\Sigma_0$, $p$, \Rd\ and $\beta$ and compute the disk SED (integrated over the
disk surface) for a distance $D$=100 pc and an inclination of 66\degree.

\begin{figure} 
\centerline{ \psfig{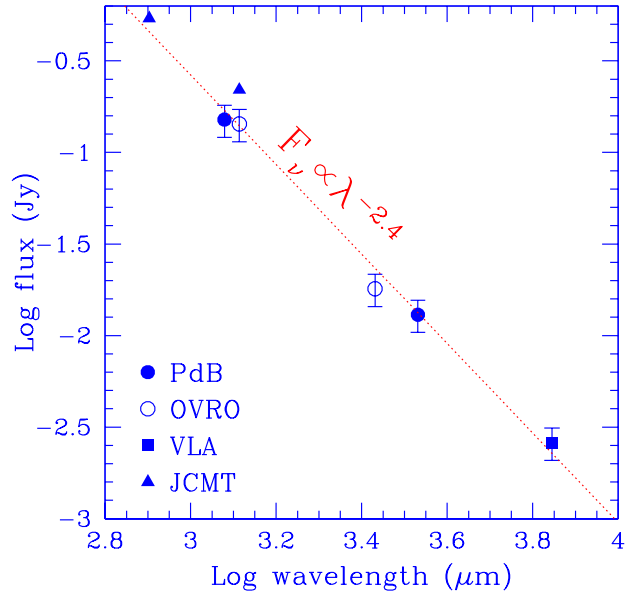} }
\caption{Observed fluxes of CQ~Tau. OVRO points are from Mannings \& Sargent
(\cite{MS97}); PdB from Dutrey, quoted by Natta et al.~(\cite{Nea01}); JCMT from
Mannings, quoted by Natta et al.~(\cite{Nea01}); the VLA 7mm point is from 
Testi et al.~(\cite{Tea01}).
Note that we have plotted for all interferometric points a calibration
 uncertainty of $\pm$20\%, even when the formal errors are smaller.
The best fitting slope -2.4 has been derived from the interferometric data
only.}
\label{mmfluxes}
\end{figure}

We first select the models that reproduce the {\it integrated}
observed fluxes at 1.3, 3 and 7mm (see Fig.~\ref{mmfluxes}) within the
observational uncertainties, keeping in mind the 3.6cm results of
Testi et al.~(\cite{Tea01}), who constrain any possible contamination from
free-free emission at millimeter wavelength to be negligible.
For the accepted models we then compute 7mm
synthetic maps and compare them to the observed one. To this purpose we
followed the method outlined by Wilner et al.~(\cite{Wea00}): each model image
is Fourier-transformed and sampled on the same ($u,v$) points as the 7mm C- and
D-array VLA datasets; the resulting visibilities are then treated in the same
way as the observational data in order to produce synthetic maps to be compared
with the observed ones. The position angle of the disk is not well constrained
by the gaussian fits of the observed emission. We  computed a set of models
with different position angles on the sky, the best results are obtained for
position angles in the range 20\degree\ to 30\degree, depending on the
other disk parameters. 

\begin{figure} 
\centerline{ \psfig{file=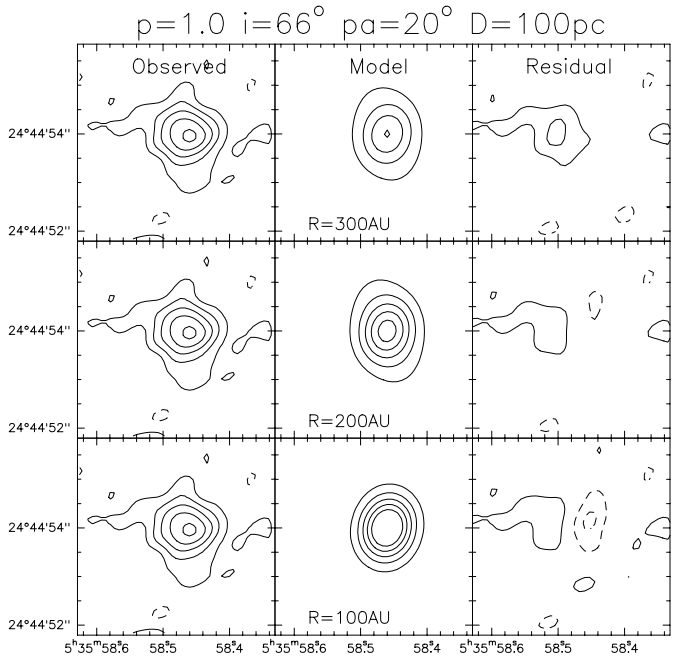,width=8.8cm,angle=-90} }
\caption{Comparison between observed and synthetic 7~mm images of the CQ~Tau
protoplanetary disk. The observed image is shown in the left panels, with the
same contour levels as in Fig.~\ref{fcqtau}: contours start at 2$\sigma$
(positive and negative) and are spaced by 2$\sigma$. The synthetic models 
images are shown in the central panels for different values of the
disk radius, the surface density power-law index (p), the inclination (i) and 
position (pa) angles are the same for all the models shown in this figure:
$p$=1.0, i=66$^\circ$, pa=20$^\circ$. The disk radii are R=100, 200, and 300~AU
as labelled. On the right panels we show the residuals after subtracting the
model images from the observed image. Contour levels are the same in all
panels.
{\it All the models shown here reproduce the observed integrated
fluxes at 1.3, 3 and 7mm.}
}
\label{maps_p10}
\end{figure}

\begin{figure} 
\centerline{ \psfig{file=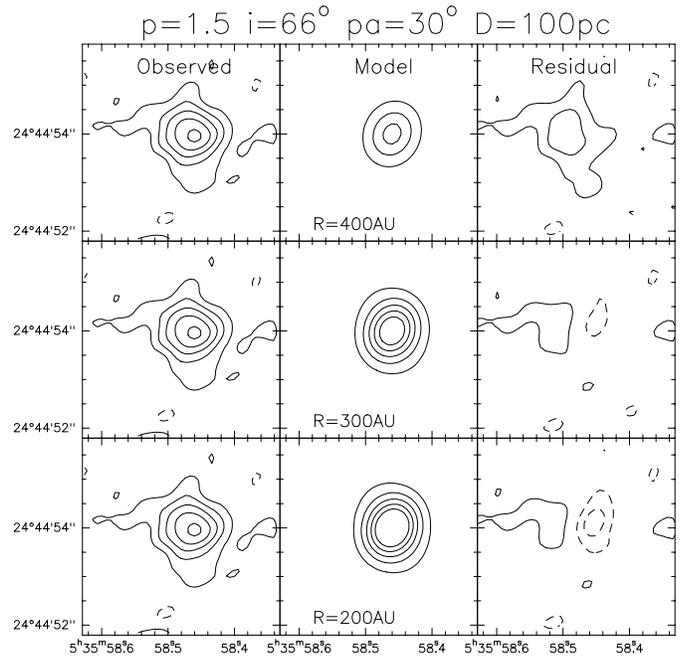,width=8.8cm,angle=-90} }
\caption{Same as Figure~\ref{maps_p10} but for $p=1.5$, pa=30$^\circ$,
and R=200, 300, and 400~AU.}
\label{maps_p15}
\end{figure}

\begin{figure} 
\centerline{ \psfig{file=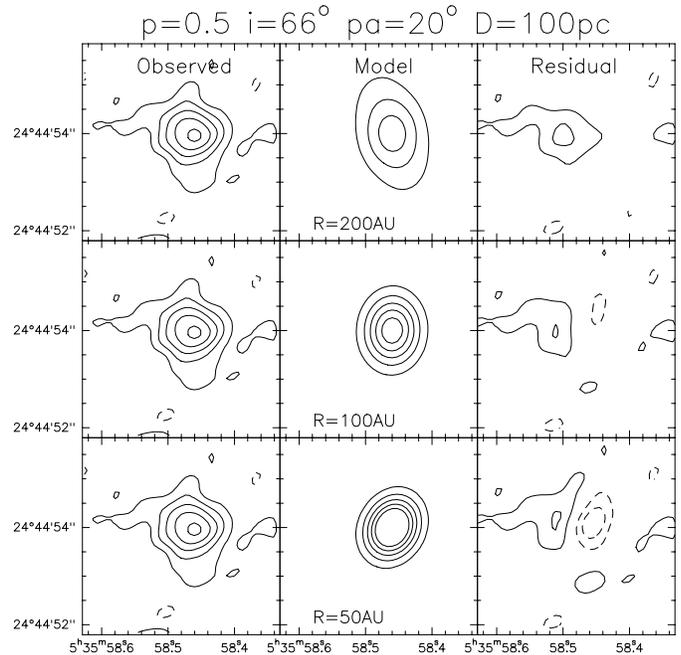,width=8.8cm,angle=-90} }
\caption{Same as Figure~\ref{maps_p10} but for $p=0.5$ and R=50, 100,
and 200~AU, respectively.}
\label{maps_p05}
\end{figure}

We  compare  the observed and model maps by
showing the residual image obtained by subtracting the noise-free model from the
observed image. 
The results for $p$=1.0 are shown in
Fig.~\ref{maps_p10}.  Since the disks are optically thin at most radii,
(except the very inner regions with R$<$8~AU) and they have to
reproduce the total integrated flux, very large disks (R$\ge$300~AU) predict an
observed surface brightness that is too low compared with observations, while
small disks (R$\le$100~AU) are too compact and have a much higher surface
brightness than observed. The best match between observations and model is
found for R$\sim$200~AU. We also investigate the possibility of shallower or
steeper surface density gradients. Steep surface density gradients ($p$=1.5,
Fig.~\ref{maps_p15}) require larger disk radii to reproduce the observed
images, the best match in this case is for R$\sim$300~AU.  Shallow surface
density gradients have been recently claimed for many TTS disks
(Kitamura et al.~\cite{Kea02}); the comparison of our CQ~Tau map with models
with $p$=0.5 is shown in Figure~\ref{maps_p05}. The best match, in this case,
is for R$\sim$100~AU, the residuals are marginally larger than in the
$p$=1.0 and 1.5 best models, but the difference is not really significant.  
The dependence of the outer radius on the surface density law
is not unexpected, see also the discussion of Mundy et al.~(\cite{Mea96}).
In all cases the outer radius is slightly or significantly larger than
the FWHM radius  derived from the 2D gaussian deconvolution of the 7mm map;
especially if $p\ge 1$, our observations are  consistent with CQ~Tau having
a rather large disk, comparable in size to the CO disks observed around
other Herbig Ae stars (Mannings \& Sargent~\cite{MS97}; \cite{MS00}).


The values of the exponent $\beta$ for the best matching models range from 0.55
for disks with flat density profile ($p$=0.5) to 0.6 for $p$=1 and 0.7 for
$p$=1.5.  In all these models, at 1.3~mm the contribution of the inner
optically thick disk ($\sim$8 AU at most) to the total flux is $<$10\%.
Thus
our modeling of the observed 7mm VLA map, together with the constraint on the
integrated fluxes from 1.3 to 7~mm, confirm that the millimeter emission of CQ
Tau originates from an optically thin disk with grains very different from
those in the ISM. No model with a significant contribution from an optically
thick inner disk fits the data, even if we relax the condition that the
CQ~Tau disk is highly inclined. We could not find any model with ISM dust in
the disk midplane that would reproduce the observations.  On the other extreme,
there is no possible fit  if the opacity is wavelength-independent
($\beta\sim$0.0), as it would be for extremely large grains (``pebbles'').

The disk mass of the best matching models does not depend strongly on $p$
or \Rd\ (due to the negligible contribution from the optically thick
region). Assuming a gas-to-dust mass ratio of 100, calculated disk masses are of 
the order of 0.1\Mstar (0.012, 0.016 and 0.017 \Msun for p=0.5, 1.0, and 1.5,
respectively). Note, however,
that the models do not constrain the actual disk mass, but rather the product
M$_{dust}\times \kappa_0$, and the values just quoted correspond to a $\kappa_0$=1
cm$^2$ g$^{-1}$ for dust particles, independently of $\beta$ (see \S4).



\section {Discussion}



If we are observing thermal, optically thin emission
from dust particles, then the derived opacity law is quite robust.
Not only is it independent of the disk geometry, but  also of
the temperature values and/or gradient, as long
as T$\gg h\nu/k$. This is always the case in all the models
we have considered.  Similarly, density
irregularities in the disk will have no effect on the measured $\beta$.
The models of the  previous section assume that the disk is flared and
that dust and gas are well mixed together. This is likely to be the case,
since both the shape of the SED at shorter wavelengths (Chiang et
al.~\cite{Cea01}; Natta et al.~\cite{Nea01}) and the characteristic
polarimetric variability of CQ~Tau (Natta \& Whitney~\cite{NW00}) are well
fitted by flared disks.  Our millimeter observations agree with this result,
but are also consistent with geometrically flat disks. In fact, because of the
optically thin emission, the millimeter images are probing the bulk
of the dust material which is located on the mid-plane in flared disks.
Nevertheless, it is important to note that flat disks models are not
self-consistent and require an ``{\it ad hoc}'' choice of the dust
temperature profile.
In this context
it is also important to note that the 3.6cm upper limit measured by Testi
et al.~(\cite{Tea01}) imply that any possible contamination of the millimeter 
flux from free-free emission must be very low ($\ll$10\% at 7mm), and its
possible impact on the derived dust opacity law is negligible.



Our observations exclude the possibility that the CQ~Tau disk is very
small ($\le$30~AU), optically thick and composed of ISM grains
(see the discussion in Testi et al.~\cite{Tea01}). If this were the case,
the 7mm emission would not be resolved and the match between the observed
and expected disk maps would be much worst than the lower panels of
Figures~\ref{maps_p10}, \ref{maps_p15}, and~\ref{maps_p05}.

Let us reiterate here that the grains we are discussing are those located in
the outer ($R\ge 5-10$~AU) disk midplane, which contains most of the mass.
We expect that grains on the disk surface will be different, very likely much
smaller.
Those ``surface" grains determine the SED at shorter wavelengths, as
discussed, for example, by CG97.
Chiang et al.~(\cite{Cea01}) obtain a  rather good
fit to the CQ~Tau SED with a face-on disk model where the surface
dust is a MRN distribution of iron and  amorphous olivine with minimum and
maximum radii of 0.01 and 1~\um, respectively. When the temperature drops
below 150~K, the silicate grains are coated with thick mantles of water ice,
whose features at 45 and 62~\um\ contribute to enhance the far-infrared flux.
The midplane dust has a maximum size of 1~mm, size distribution 
$n(a) \propto a^{-3.5}$;  $\Sigma$ is  $\propto R^{-1.5}$, the disk model
has a mass of 0.04~\Msun\ and  outer radius  180~AU. These disk parameters
are not very different from what we derive. However, a maximum grain size of
1~mm will  result in an opacity which drops too steeply at 7~mm
to account for our observations. Natta et al.~\cite{Nea01} fit
the same SED (which extended only to 2.6~mm) with a small (\Rd=50~AU), 
very tilted ($\theta$=66~deg) disk, same surface density profile,
\Md=0.02~\Msun\, a mixture of silicates and
graphite for the surface dust similar to what adopted here, and midplane dust
opacity $\kappa\propto \lambda^{-1}$. Also this set of parameters
would fail to reproduce the characteristics of our 7~mm emission.
As noticed in \S~3, the disk emission at millimeter wavelengths is determined
by the midplane grain and disk properties, and it is not affected by the nature
of the surface dust. Consequently, in this paper, we have not made any effort
to select the best parameters for the surface dust to fit the
near and far-infrared CQ~Tau SED.

The Chiang et al.~\cite{Cea01} and Natta et al.~\cite{Nea01} 
results show that disk models,
that allow for different dust properties on the disk surface and midplane,
can account for the CQ~Tau SED at all wavelengths. This is important, because a
serious concern in our analysis is
the  possibility  that CQ~Tau retains a residual envelope of dust which
enshrouds the disk and dominates its millimeter emission. 
The effects of such an envelope on the disk temperature and
on the overall observed SED have been discussed by Natta~(\cite{N93}) and more
recently, among others, by Kikuchi et al.~(\cite{Kiea02}). Although no model
specific for CQ~Tau has been computed, envelopes with shallow density profiles
(shallower than $r^{-1}$) produce mid and far-infrared excess emission in
the SED, not very different from what is observed in CQ~Tau (Natta et
al.~\cite{Nea01}).  However, as we have seen, also two-layer disk models can
account for it. Additionally, the contribution of 
envelopes to the small-scale millimeter flux detected with the interferometers
is generally negligible (see, for example, Butner et al.~\cite{Bea94}). This
is very likely the case of CQ~Tau, given that the JCMT flux at 1.3mm, obtained
with a beam of about 20 arcsec, is only $\sim$50\% larger than the flux
measured by the interferometers, and that this difference is in fact comparable
with the calibration uncertainties of both measurements.

Let us, therefore, consider the implications of our result, namely that the
mean opacity of the grains in the CQ~Tau outer disk midplane has a wavelength
dependence $\kappa\propto \lambda^{-0.6\pm 0.1}$ in the millimeter. It is
customary to interpret  small values of $\beta$ as evidence of grain growth.
Miyake \& Nakagawa~(\cite{MN93}) have shown that the  mass opacity coefficient
of a distribution of silicates and ice with $n(a)\propto a^{-3.5}$ has $\beta
\simless 1$ 	in the millimeter if the maximum grain radius increases to
values of 0.1-10 cm,
depending on the porosity of the grains.  
Recently, D'Alessio et al.~(\cite{DAea01}) have explored
the effect of grain growth on disk properties using the dust model of Pollack
et al.~(\cite{Pea94}).  They find that $\beta \sim 0.6$ is reproduced for $T<100$K,
i.e., when ices are present, by a size distribution $n(a)\propto a^{-2.5}$ with
minimum and maximum grain radius of 0.005 \um\ and 5 cm, respectively.  If this
is indeed the case, one needs to revise upward the estimated disk mass of CQ~Tau
by a factor of about 6, to compensate for the corresponding decrease of the
opacity (D'Alessio et al.~\cite{DAea01}).  The disk mass is then 
approximately 6\% of the stellar mass.  This is a rather large value, but not
exceptionally so (Natta et al.~\cite{NGM00}).

Calvet et al.~(\cite{Cea02}) have used the D'Alessio et al.~(\cite{DAea01})
models to investigate the properties of the T Tauri star TW~Hya, which has a
millimeter spectral index $\alpha\sim$2.4, practically identical to that of
CQ~Tau,  and is also resolved at 7mm (Wilner et al.~\cite{Wea00}).  They find
that the grains in the outer disk of TW~Hya must have grown to sizes of $\sim
1$cm, and that the disk has a mass of about 0.06 \Msun, i.e., 10\% of the
stellar mass. The profile of scattered light at 1.1 and 1.6 \um\ indicates that the disk is flared, rather than geometrically thin (Weinberger et al.~\cite{Wea02}).

The similarities between \CQ\ and TW~Hya are striking. Both objects
seem to have flared disks, where gas and dust have to be well
mixed; grains have certainly been processed, and the original size distribution
has been significantly altered, but their outer disks retain the main
features of youth.  Models of the evolution of grains in the solar nebula
predict a first phase where grains grow and simultaneously settle on the disk
midplane by collisional coagulation, controlled by the drag of the gas in the
disk (see Weidenschilling~\cite{W00} and references therein).  The typical
outcome (Weidenshilling~\cite{W97}) is a concentration of solids in the
midplane, with about 1/2--1/3 of the original dust mass in large bodies
(diameter $>$1 km) and a tail of smaller bodies distributed  between the
initial size (1\um) and a maximum of about few metres.  A small fraction of the
grains may be easily stirred by turbulence to higher altitudes than predicted,
providing the opacity (to the stellar radiation) necessary to keep the disk
flared.  Given the many uncertainties, these results are at least qualitatively
consistent with the analysis of millimeter data for CQ Tau and TW Hya.

A problematic aspec is that both CQ Tau and TW Hya are rather ``old''
pre-main-sequence stars, with estimated ages of about 10 Myr. The timescale of
the collisional coagulation process is typically a few thousands times the
Keplerian period, i.e., $\simless 10^6$ yr at 100 AU from CQ~Tau.  After this
time, one can expect further evolution to occur, clearing the gaseous disk and
depleting further the population of small solids (Lagrange et al.~\cite{Lea00}
and references therein).  It is possible that in the outer disk collisional
coagulation proceeds much more slowly than currently estimated, due, e.g., to
slow-decaying turbulence or to a sticking probability much lower than  assumed
(Beckwith et al.~\cite{Bea00} and references therein).  Until calculations that
explore these (or other) possibilities are performed, it will be difficult to
understand the exact implications of the observational results on grain growth
in the outer disks of pre-main--sequence stars that are now becoming available.


\section{ Conclusions}

We present new subarcsecond resolution 7mm observations that clearly
resolve the dusty disk surrounding the 10 Myr old 1.5 M$_\odot$
pre-main-sequence
star CQ Tauri, and we investigate millimeter observations from 1.3mm to 7mm
using a two-layer flared disk model. The main conclusions are:

\begin{enumerate}

\item The 7mm emission is resolved in approximately the north-south
direction, with aspect ratio that implies an inclination of about
70 degrees, in good agreement with that derived from polarimetric
and photometric variability (Natta \& Whitney~\cite{NW00}).

\item The resolved size of the 7mm emission rules out the possibility
that the millimeter spectral index of $\sim2.4$ can be accounted for
by optically thick emission.

\item The comparison of the millimeter data with disk models shows that
the mean dust opacity in the outer disk has a wavelength dependence
$\kappa \propto \lambda^{-0.6\pm0.1}$.
The observations are not consistent with a steeper power law dependence,
such as found in ISM dust, nor with a grey opacity, as expected if the
particles were much larger than the wavelengths of the observations.
This result does not depend critically on any of the model parameters,
since the dust emission is almost entirely optically thin.

\item The most likely interpretation of the shallow millimeter spectral
index is that the dust particles have grown to sizes much larger than
the sub-micron ones typical of the ISM. A mixture of grains of different
sizes and composition with
$n(a)\propto a^{-2.5}$ and maximum size of few centimeters can reproduce the
observation (Miyake \& Nakagawa~\cite{MN93}; D'Alessio et al.~\cite{DAea01}).
The evidence for grain growth and evolution is clear.

\item The CQ Tau millimeter opacity law is similar to that in TW~Hya (Calvet et
al.~\cite{Cea02}).  Both  CQ~Tau and TW~Hya have ages of about 10 Myr, and  one
expects that their disks have evolved significantly. In fact, it is
somewhat surprising to find that they have not evolved more, as predicted by
current models of disk evolution and planet formation (Ruden~\cite{R99}).  It
is clear that calculations that explore more extensively a larger parameter
space are  required.  Here we can only stress  that, at present, spatially
resolved multiwavelengths millimeter observations provide  the best
observational constraint for models of the early stages of the evolution of the
dusty disk.

\end{enumerate}


\begin{acknowledgements}
We have greatly profited from discussions with several of our collegues,
among them C. Dominik, F. Marzari, and S. Weidenschilling.
This work was partly supported by  ASI grant  ARS 1/R/27/00 and ARS-1/R/073/01
to the Osservatorio di Arcetri.
This research was partially supported by NASA Origins of Solar Systems
Program grant NAG5-8195.
\end{acknowledgements}

\end{document}